\documentclass[11pt,twoside]{article}

\usepackage{asp2006}
\usepackage{epsf}
\usepackage{psfig}
\usepackage{lscape}
\usepackage{graphicx}

\newcounter{dddr}

\begin{document}
\title{The far-UV break in quasar energy distributions: dust?}   

\author{Luc Binette,  Yair Krongold and Sinhue Haro-Corzo
\affil{Instituto de Astronom\'{\i}a, Universidad Nacional Aut\'onoma de
M\'exico, Apartado Postal 70-264, 04510 M\'exico, DF, Mexico}}
\author{Anja Andersen \affil{Dark Cosmology Center, Juliane Maries Vej 30, DK-2100 Copenhagen, Denmark}}

\begin{abstract}
A prominent continuum steepening is observed in quasar energy
distributions near $1100$\AA. We review possible interpretations for
the origin of the so-called far-UV break, putting emphasis on those
that favor the emergence of an upturn in the extreme-UV.
\end{abstract}


\section{Introduction}

The spectra of quasars and Seyfert galaxies show strong emission
lines superimposed onto a bright continuum.  The continuum contains
a significant feature in the optical-ultraviolet region, known as
``the Big Blue Bump''.  Due to the huge photoelectric opacity of the
Galaxy, the Spectral Energy Distribution (SED) of the ionizing
continuum, between the Lyman limit and the soft X-rays (EUV--X), is
hardly known. Fortunately, owing to the redshift effect and the
transparency of the quasar environment, it has been possible to
infer the SED of quasars down to $\sim$350\AA. Composite energy
distributions were derived by Zheng et\,al. (1997) and later by
Telfer et\,al. (2002) using archived HST-FOS spectra. Before
averaging, each spectrum was dereddened for Galactic absorption as
well as statistically corrected for the absorption due to
intergalactic Ly$\alpha$ absorbers and Lyman limit systems.

A striking feature of the composite quasar SED is that a significant
steepening occurs around 1100\AA, leading to a far-UV powerlaw of
index $\nu^{-1.7}$ ($F_{\nu}\propto \nu^{+\alpha}$). Korista,
Ferland \& Baldwin (1997) pointed out the difficulties of
reproducing the equivalent widths of the high ionization lines of
He{\sc ii} (1640\AA), C{\sc iv} (1549\AA) and O{\sc vi} (1035\AA),
assuming a powerlaw as soft as $\nu^{-2}$. State of the art
photoionization models favor a much harder SED, one that peaks in
the extreme-UV beyond 22\,eV (e.g. Casebeer, Leighly \& Baron 2006;
Korista et\,al. 1997; Baldwin et\,al. 1995). The far-UV break is
clearly seen in individual spectra (see Binette et\,al. 2005:
hereafter B05). The amount of steepening varies considerably from
object to object. Binette \& Krongold (2006) recently analyzed the
spectrum of Ton\,34 ($z=1.928$), which is the object with the
steepest break known, with a far-UV behavior given by $\nu^{-5.3}$.
However, strong emission lines of OVI and CIV  are present. There is
no generally accepted interpretation of the nature of the far-UV
break. We review below possible absorption mechanisms that would
give rise to the break and at the same time allow the emergence of
an upturn in the extreme-UV, in order that the ionizing SED  be as
hard as needed in photoionization calculations of the emission
lines.


\section{Possible causes for the UV-break}

We hereafter assume that the far-UV break results from absorption
and  will consider two possibilities: (I)-- H{\sc i} [Ly$\alpha$,
$\beta$, $\gamma$ ... and bound-free] and (II)-- interstellar dust.
We will consider four locations for the absorbing medium: ($i$)
intergalactic, ($ii$) local to the quasar ISM, ($iii$) accretion
disk photosphere and ($iv$) accelerated outflow from quasars. The
resulting eight cases are illustrated in Fig.\,1 with labels 1--4
for H{\sc i} and A--D for the dust. The aim is to resolve the
problem of the ionizing SED, which appears to be too soft.  Among
the eight cases reviewed, some have the potential to resolve the
softness problem, either because the local BELR sees a different SED
(intergalactic absorption), or because the absorption in the UV is
followed by  a flux upturn at higher energies.

\begin{figure}[!ht]
\begin{center}
\includegraphics[width=9.5cm,keepaspectratio=true]{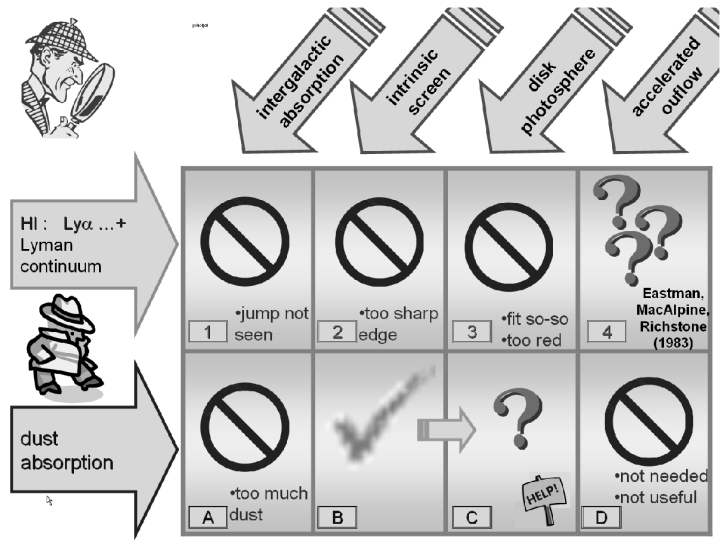}
\caption{Diagram  illustrating the 8 cases due to either H{\sc i} or
dust absorption. Barred circles denote rejected or unpromising
cases.}
\end{center}
\end{figure}

\begin{list}{Case~\arabic{dddr} --}{\usecounter{dddr}\setcounter{dddr}{0}}
\item Binette et\,al. (2003) studied this possibility. They assumed
an H{\sc i} behavior (as a function of $z$) proportional to the gas
density expected from the warm-hot intergalactic medium. Although
they could reproduce the TZ02 composite, they rejected this
possibility, since their model predicted a jump in flux blueward of
1216\AA\ (observer-frame) that is not observed.

\item An absorber at the quasar redshift results in a sharp
absorption edge at 912\AA\ as well as in an Ly$\alpha$ absorption
line. Such pronounced and sharp features do not match the far-UV
steepening discussed above.

\item State of the art `naked' accretion disk (NAD)
models predict a steepening (i.e. Lyman edge) near Ly$\alpha$.
However, the far-UV break is not reproduced well by NAD models (c.f.
Fig.\,22 in Hubeny et\,al. 2000). The Lyman edge from a NAD model is
not followed by a flux upturn at higher energies, hence the softness
problem remains unresolved. The same may be said of the comptonized
accretion disk model although in that case the UV break is well
reproduced (Zheng et\,al. 1997).

\item Eastman, MacAlpine \& Richstone (1983) could generate a steepening of the
continuum by having absorptions clouds progressively accelerated up
to 0.8$c$. Exploration of a different behavior of the H{\sc i}
opacity with velocity would be welcomed, as these calculations might
shift the break position to the observed value near 1100\AA\
(instead of 1216\AA), and possibly produce a flux upturn beyond
20\,eV. Furthermore, by including He{\sc i} opacity, such models
might explain the curious dip observed at 500\AA\ in one of the most
studied quasar, HE\,2347$-$4342 (see Fig.\,2). Interestingly, both
the 1100\AA\ and 500\AA\ breaks are blueshifted by comparable
amounts with respect to rest-frame He{\sc i} and H{\sc i} Ly$\alpha$
(at 584\AA\ and 1216\AA, respectively).

\end{list}

\begin{figure}[!ht]
\plottwo{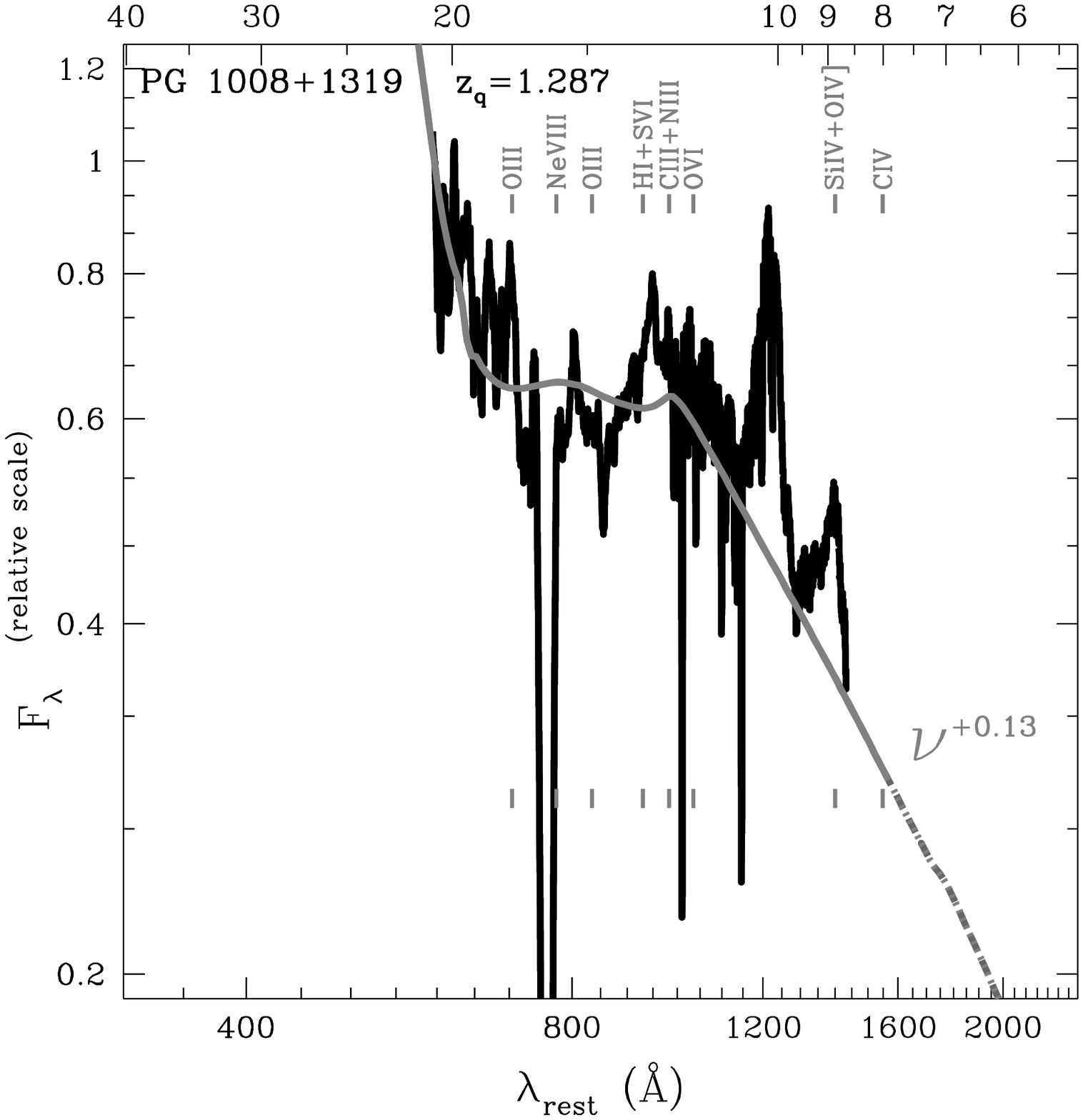}{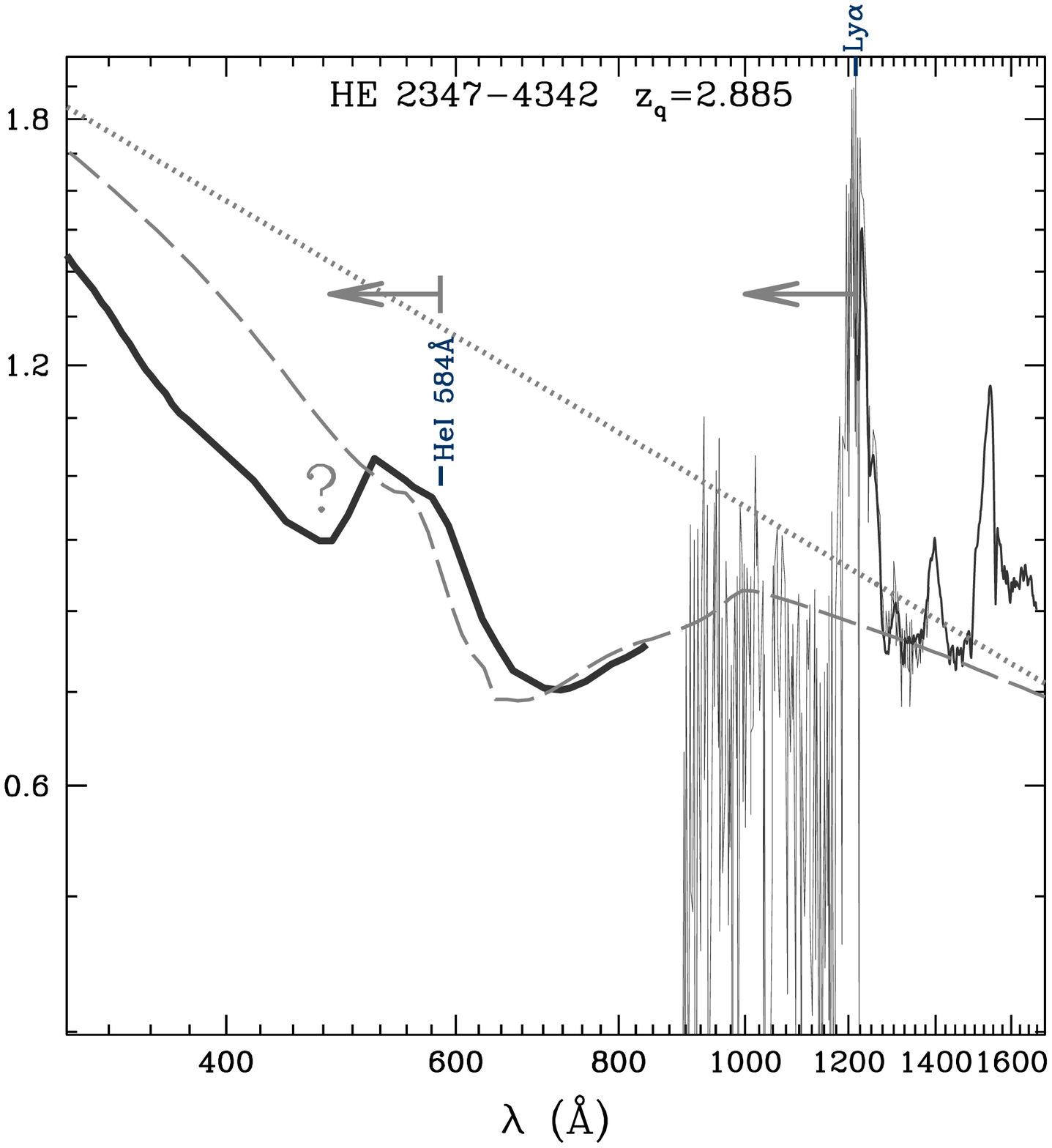} \caption{Left panel:
rest-frame spectrum of PG\,1008+1319 (black line). The  silver line
represents an absorption model assuming extinction by cubic diamonds
and an intrinsic SED consisting of a powerlaw (dotted line).  Right
panel: spectrum of HE\,2347$-$4342 (thick and thin black lines)
extracted from Reimers et\,al. (1998). A powerlaw SED (dotted line)
and a dust absorption model (dashed line) are also shown. }
\end{figure}

\begin{list}{Case~\Alph{dddr} --}{\usecounter{dddr}\setcounter{dddr}{0}}

\item  B05 studied the possibility of intergalactic
dust consisting of nanodiamond grains (other dust compositions were
unsuccessful).  They could reproduce the dip displayed by the far-UV
indices when plotted as a function of redshift.  This possibility
was rejected, since it required too much intergalactic dust
($\sim$17\% of the cosmic carbon). It also required  that  only
nanodiamond dust is formed.

\item Assuming a nanodiamond dust component intrinsic to each quasar,
B05 successfully reproduced the far-UV break observed in individual
quasars of the TZ02 sample.  The flux upturn taking place below
650\AA\ in these models near allow the intrinsic SED to be much
harder than indicated by extrapolating of the flux near the UV
break. Such an upturn was identified in 4 quasar spectra.  An
example, PG\,1008+1319, is shown in the left panel of in Fig.\,1.
From a spectrum that combines various archives, Binette \& Krongold
(2006) reported a far-UV rise in Ton\,34. A 6th example is provided
by HE\,2347$-$4342 (see upturn shortward of 700\AA\ in Fig.\,2).
Shang et\,al. (2005) explored the possibility of ISM and SMC-like
extinction.

\item It would be interesting to compute the SED emerging from a NAD photosphere
that contains small amounts of nanodiamond dust.  If the break is
accounted for by dust, much harder (hotter) disk SEDs could be
envisaged. Nanodiamonds have so far been identified in emission
around 3 stellar disks. A large UV fluence may facilitate dust
formation.

\item We consider the hypothesis of an accelerating dust outflow not to be needed nor useful
for the purpose of explaining the far-UV break.

\end{list}

\section{Conclusions}

We find that the three cases 4, B, and C are worth exploring
further, as they may reconcile the observed downturn in the far-UV
with the need of a harder SED to account for the high ionization
emission lines. The exploration of absorption by an intrinsic
crystalline carbon dust screen (case C) is at this stage the most
developed hypothesis. We are currently exploring its consequences in
the infrared and UV in order to provide ways to falsify such models.

\acknowledgements This work was funded by the CONACyT grants J-49594
and J-50296, and the UNAM PAPIIT grant IN118905. The Dark Cosmology
Centre is funded by the Danish National Research Foundation.


\end{document}